\documentclass[doublecol]{epl2}
\usepackage{graphicx}
\usepackage{subfigure}

\title{Single nonmagnetic impurity resonance in FeSe-based 122-type superconductors as a probe for pairing symmetry}

\author{Qian-En Wang, Zi-Jian Yao and Fu-Chun Zhang}

\institute{Department of Physics and Center of Theoretical and
Computational Physics, The University of Hong Kong, Hong Kong,
China}




\abstract{We study the effect of a single non-magnetic impurity in
A$_{y}$Fe$_{2-x}$Se$_{2}$ (A=K, Rb, or Cs) superconductors by
considering various pairing states based on a three-orbital model
consistent with the photoemission experiments. The local density of
states on and near the impurity site has been calculated by solving
the Bogoliubov-de Gennes equations self-consistently. The
impurity-induced in-gap bound states are found only for attractive
impurity scattering potential, as in the cases of doping of Co or
Ni, which is characterized by the strong particle-hole asymmetry, in
the nodeless $d_{x^2-y^2}$ wave pairing state. This property may be
used to probe the pairing symmetry of FeSe-based 122-type
superconductors.}

\begin{document}

\maketitle

\section{Introduction}

The FeSe-based 122-type A$_{y}$Fe$_{2-x}$Se$_{2}$ (A=K, Rb, or Cs)
high-$T_c$ superconductors have been the focus of intensive
attention due to their distinctive high transition
temperatures($\sim$30K) \cite{J.Guo,M.H.Fang,R.H.Liu,A.K-M} and
novel Fermi surface topology. Angle-resolved photoemission
spectroscopy (ARPES) \cite{Y.Zhang,D.Mou,X.P.Wang,T.Qian,L.Zhao} and
density functional theory (DFT) calculations
\cite{C.Chao,L.Zhang,X.W.Yan,I.R.Shein} show that there is no hole
pocket around the $\Gamma$ point of the Brillouin zone (BZ), and
large electron pockets are observed around the M point which is
located at the corners of the folded BZ \cite{Y.Zhang}. Effort has
been made to investigate the pairing symmetry of the Fe-based
superconductors. The popular $s_{\pm}$ pairing symmetry in iron
pnictide 1111-type superconductors in which the hole pockets are at
the $\Gamma$ point, has been predicted by spin-fluctuation theory
\cite{K.Kuroki, I.I.Mazin}. However, for FeSe-based 122-type
superconductors, due to the absence of the hole pockets, nodeless
superconducting (SC) gap can be originated from various pairing
symmetries. Spin-fluctuation theory predicted $d_{x^2-y^2}$ wave
pairing state \cite{T.Das,T.A.Maier,F.Wang,T.Saito}, while $s$ wave
pairing state was argued to be more favorable against strong
disorder \cite {Y.Zhou}. Moreover, a co-existence candidate of
anisotropic $s$ and $d_{x^2-y^2}$ wave pairing symmetries has also
been predicted \cite{R.Yu}. These results reveal the competitive
nature of the pairing state in these novel superconductors.

The response of a SC system to impurities can be a probe for the SC
pairing symmetry \cite{A.V.Balatsky}. The local electronic states
can evidently be influenced by impurities. In a multi-band SC
system, the SC order parameter may or may not be sensitive to the
existence of impurity \cite{J.M.Byers}. Doping of Co and Zn in iron
pnictide 1111-type superconductors has been realized by several
experimental groups
\cite{A.Kawabata1,A.Kawabata2,A.S.Sefat,A.L-J,Y.K.Li}.
Theoretically, the effects of impurity in iron pnictide
superconductors have also been investigated intensively
\cite{T.Zhou1,M.Matsumoto,T.Zhou2,Y.Senga,D.G.Zhang,T.Kariyado,W.F.Tsai,Alireza1,Alireza2}.
The calculated results of $T$-matrix method based on a four-band
model show the in-gap impurity resonance states \cite{D.G.Zhang} for
$s_{\pm}$ wave pairing symmetry, and results obtained by solving the
Bogoliubov-de Gennes (BdG) equations in the framework of a detailed
five-orbital model also demonstrate that the $s_{\pm}$ wave pairing
symmetry is characterized by a bound state in the vicinity of the
Fermi level in the presence of a single nonmagnetic impurity
\cite{T.Kariyado}. However, for FeSe-based 122-type superconductors,
the impurity resonance bound state, which is located away from the
Fermi level, were obtained only in $d_{x^2-y^2}$ wave pairing state
\cite{J.X.Zhu}.

In this work, we use self-consistent BdG equations to study the
effect of a single nonmagnetic impurity in FeSe-based 122-type
superconductors modeled by a three orbital Hamiltonian. The local
density of states(LDOS) and the real-space SC order parameters are
calculated. It has been reported that the effective on-site impurity
potentials can be repulsive or attractive which is related to the
energy difference between the impurity 3d levels and Fe 3d level for
various impurity atoms. Realistic estimation of the impurity
potential in iron pnictide superconductors predicts that the
potential difference for Zn, Co, and Ni are negative, whereas for Mn
it is positive \cite{K.Nakamura}. Therefore both repulsive(positive)
and attractive(negative) scattering potentials have been considered
in our calculation which may simulate different dopants, such as Zn,
Co, Ni, and Mn etc.. Responses of isotropic $s$, anisotropic $s$,
and $d_{x^2-y^2}$ wave pairing states to impurity are examined,
respectively.

\section{Model and Methodology}

\begin{figure}[t]
\begin{center}
\includegraphics[width=140pt]{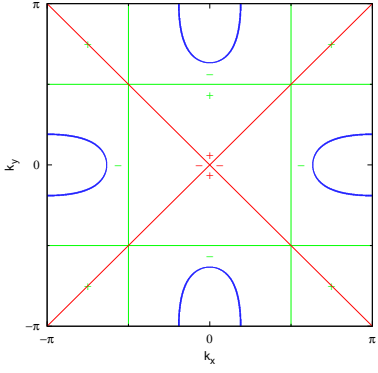}
\end{center}
\caption{(color online) The Fermi surface in the extended BZ. The
nodal lines of the gap function for anisotropic $s$ and
d$_{x^2-y^2}$ wave pairing symmetry are plotted with green and red
lines, respectively.} \label{fig1}
\end{figure}

We describe the SC system with a single nonmagnetic impurity by a
three-orbital Hamiltonian
\begin{equation}
H=H_{0}+H_{pair}+H_{imp}
\end{equation}
and the hopping, pairing, and impurity terms read
\begin{eqnarray}
&H_{0}&=\sum_{i,j,\alpha,\beta,\sigma}t_{ij}^{\alpha\beta}a_{i\alpha\sigma}^{\dag}a_{j\beta\sigma}
-\mu\sum_{i,\alpha,\sigma}a_{i\alpha\sigma}^{\dag}a_{i\alpha\sigma}\nonumber
\\
&H_{pair}&=\sum_{i,j,\alpha,\beta}(\Delta_{ij}^{\alpha\beta}a_{i\alpha\uparrow}^{\dag}a_{j\beta\downarrow}^{\dag}+h.c.)\nonumber
\\
&H_{imp}&=U_{s}\sum_{\alpha,\sigma}a_{l\alpha\sigma}^{\dag}a_{l\alpha\sigma}
\end{eqnarray}
where $a_{i\alpha\sigma}^{\dag}$($a_{i\alpha\sigma}$) denotes the
creation(annihilation) operator of electrons with spin
$\sigma=\uparrow,\downarrow$ and orbital $\alpha$ at site $i$.
$t_{ij}^{\alpha\beta}$ are the hopping integrals and $\mu$ is the
chemical potential. A single nonmagnetic impurity is located at site
$l$. Results of DFT calculation demonstrate that the electronic
structure in the vicinity of the Fermi level is dominated by
$d_{xz}$, $d_{yz}$, and $d_{xy}$ orbitals \cite{I.R.Shein},
therefore it is feasible to describe the band structure by
introducing an effective three-orbital model\cite{C.Fang}. The
independent parameters of the hopping model (in Unit: eV) are
$t^{11/22}_{i,i}$=0, $t^{33}_{i,i}$=0.4,
$t^{11}_{i,i\pm\hat{x}}$=0.05, $t^{11}_{i,i\pm\hat{y}}$=0.01,
$t^{11/22}_{i,i\pm\hat{x}\pm\hat{y}}$=0.02,
$t^{12/21}_{i,i+\hat{x}+\hat{y}}$=0.01,
$t^{13}_{i,i+\hat{x}}$=$t^{23}_{i,i+\hat{y}}$=-0.2,
$t^{13}_{i,i+\hat{x}\pm\hat{y}}$=$t^{23}_{i,i\pm\hat{x}+\hat{y}}$=0.1,
where $\hat{x}$($\hat{y}$) denote the unit vectors along $x$($y$)
directions of the lattice coordinate. Fig. \ref{fig1} shows the
heavily electron-doped Fermi surface obtained with a chemical
potential $\mu$=0.312 eV corresponding to a filling factor $n$=4.23
and it turns out that the bandwidth of the model is 2.83 eV. The SC
gap function stemming from the mean-field decoupling of the pair
scattering process
$V_{ij}^{\alpha\beta}a_{i\alpha\uparrow}^{\dag}a_{j\beta\downarrow}^{\dag}a_{j\beta\downarrow}a_{i\alpha\uparrow}$
is expressed as
\begin{eqnarray}
\Delta_{ij}^{\alpha\beta}=V_{ij}^{\alpha\beta}\langle
a_{j\beta\downarrow}a_{i\alpha\uparrow}\rangle
\end{eqnarray}
The diagonal condition of the Hamiltonian is the BdG equation
\begin{equation}
\sum_{j,\beta}\left(
               \begin{array}{cc}
                 h_{ij}^{\alpha\beta} & \Delta_{ij}^{\alpha\beta} \\
                 \Delta_{ij}^{\alpha\beta\ast} & -h_{ij}^{\alpha\beta\ast} \\
               \end{array}
             \right)\left(
                      \begin{array}{c}
                        u_{j\beta}^n \\
                        v_{j\beta}^n \\
                      \end{array}
                    \right)=\epsilon_{n}\left(
                                          \begin{array}{c}
                                            u_{i\alpha}^n \\
                                            v_{i\alpha}^n \\
                                          \end{array}
                                        \right)
\end{equation}
where
\begin{equation}
h_{ij}^{\alpha\beta}=t_{ij}^{\alpha\beta}-\mu\delta_{ij}\delta_{\alpha\beta}+U_{s}\delta_{il}\delta_{jl}\delta_{\alpha\beta}
\nonumber
\end{equation}
and the gap equation is represented as
\begin{equation}
\Delta_{ij}^{\alpha\beta}=-\frac{V_{ij}^{\alpha\beta}}{2}\sum_{n}(u_{i\alpha}^{n}v_{j\beta}^{n\ast}
+u_{j\beta}^{n}v_{i\alpha}^{n\ast})\tanh(\frac{\epsilon_n}{2k_{B}T})
\end{equation}

Single static impurity in a uniform superconductor which manifests
itself as a point-like imperfection can be regarded as a localized
orbital-independent on-site scattering potential
$U_{s}\delta_{\hat{R_l}{R_0}}$ with the impurity at $R_0$ in orbital
representation, provided that the Coulomb interaction is screened at
length scales comparable to the lattice spacing \cite{A.V.Balatsky}.
Both positive and negative scattering potential $U_{s}$ are
considered in our calculation, and by checking the results obtained
from several different values, it has been found that $U_{s}=\pm6.0$
eV can be regarded as the unitary limit.

The absence of the hole pockets of the FeSe-based 122-type Fermi
surface around $\Gamma$ point eliminates the possibility of nodes.
Calculation pertaining to the magnetic exchange couplings reveals
that the leading pairing is originated from the intra-orbital
pairing, whereas the inter-orbital components are found to be
significantly small\cite{C.Fang}. Consequently, the $s$ and
$d_{x^2-y^2}$ wave pairing symmetries are considered for
intra-orbital pairing only in order to investigate their responses
to a single nonmagnetic impurity scattering. For isotropic $s$ wave
pairing,
\begin{eqnarray}
V_{ij}^{\alpha\alpha}&=&-g_{0}\delta_{ij}
\end{eqnarray}
for anisotropic $s$ wave pairing,
\begin{eqnarray}
\nonumber
V_{ij}^{\alpha\alpha}&=&-\frac{g_{1}}{4}(\delta_{i+\hat{x}+\hat{y},j}+\delta_{i-\hat{x}+\hat{y},j}\\
&+&\delta_{i-\hat{x}-\hat{y},j}+\delta_{i+\hat{x}-\hat{y},j})
\end{eqnarray}
and for $d_{x^2-y^2}$ wave pairing,
\begin{eqnarray}
\nonumber
V_{ij}^{\alpha\alpha}&=&-\frac{g_2}{2}(\delta_{i+\hat{x},j}+\delta_{i+\hat{y},j}\\
&+&\delta_{i-\hat{x},j}+\delta_{i-\hat{y},j})
\end{eqnarray}
where $g_{0,1,2}$ are pairing amplitudes for each pairing symmetry.
The pairing patterns in real space result in specific pairing
symmetries of the gap function in k-space as for isotropic $s$ wave
channel $\Delta^{\alpha\alpha}(k)\sim\Delta_{0}$, for anisotropic
$s$ wave channel $\Delta^{\alpha\alpha}(k)\sim \cos(k_x)\cos(k_y)$,
and for $d_{x^2-y^2}$ wave channel
$\Delta^{\alpha\alpha}(k)\sim\cos(k_x)-\cos(k_y)$ in the extended
BZ, respectively. The pairing amplitudes are taken to be $g_0=0.3$
eV, $g_1=1.0$ eV, and $g_2=0.5$ eV which guarantees that the SC
coherence length $\xi \leq 4a$, where $a$ is the lattice constant.

The energy spectrum of the quasiparticle i.e., the LDOS at site $i$
is calculated via
\begin{equation}
\rho_{i}(\epsilon)=\frac{1}{N_{orb}}\sum_{n,\alpha}[|u_{i\alpha}^{n}|^2\delta(\epsilon-\epsilon_{n})
+|v_{i\alpha}^{n}|^2\delta(\epsilon+\epsilon_{n})]
\end{equation}
where the prefactor comes from taking the average of three orbitals.
The Lorentzian smearing method is used to visualize the LDOS with a
broadening width $\sigma=0.001$. All the self-consistent
calculations are performed on a 20$\times$20 lattice with a periodic
boundary condition at temperature T$=0.1$K, and the single impurity
is located at the center of the lattice.

\section{Results and Discussion}

\begin{figure}[t]
\begin{center}
\subfigure[]{\includegraphics[width=120pt]{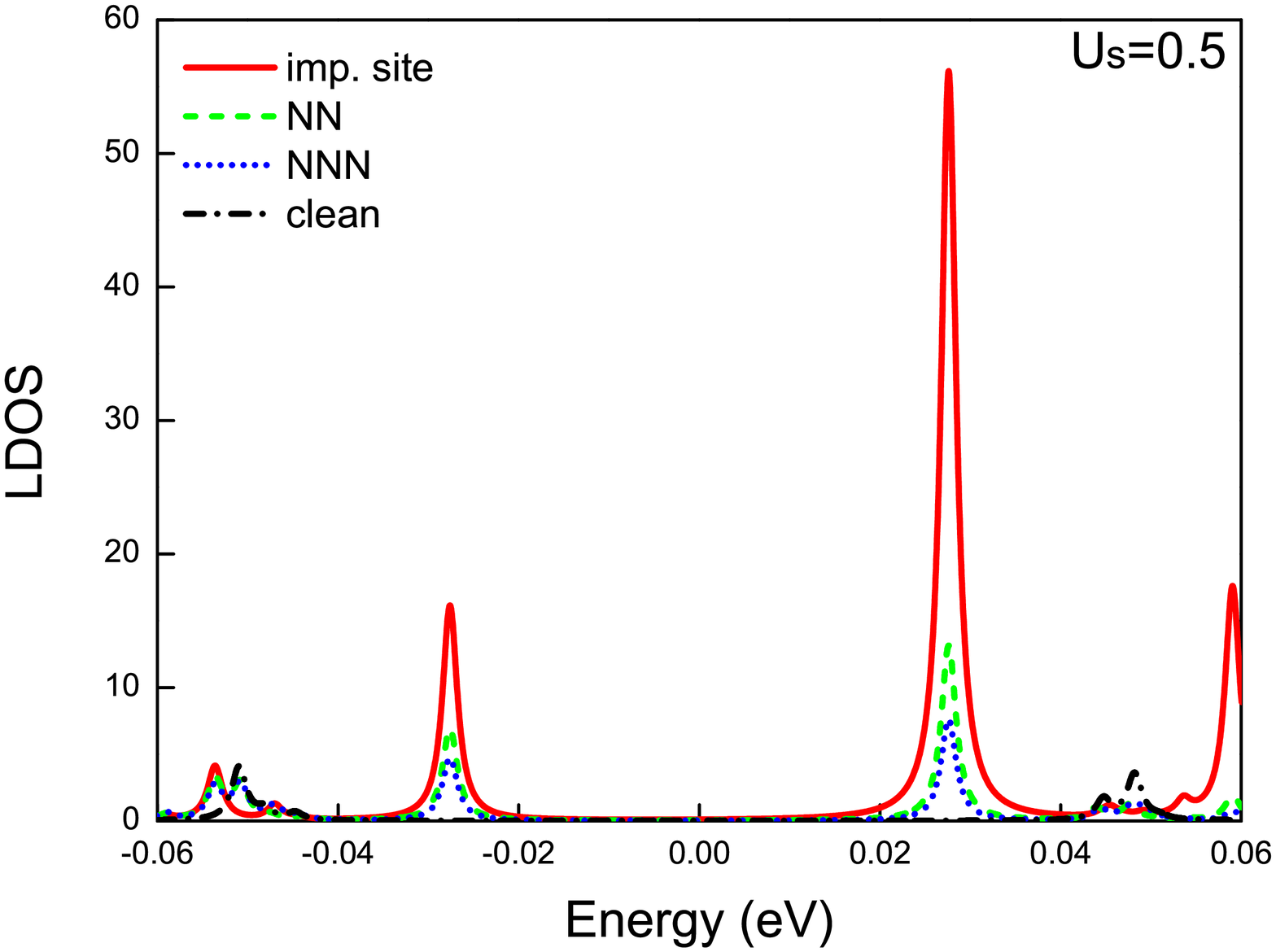}}
\subfigure[]{\includegraphics[width=120pt]{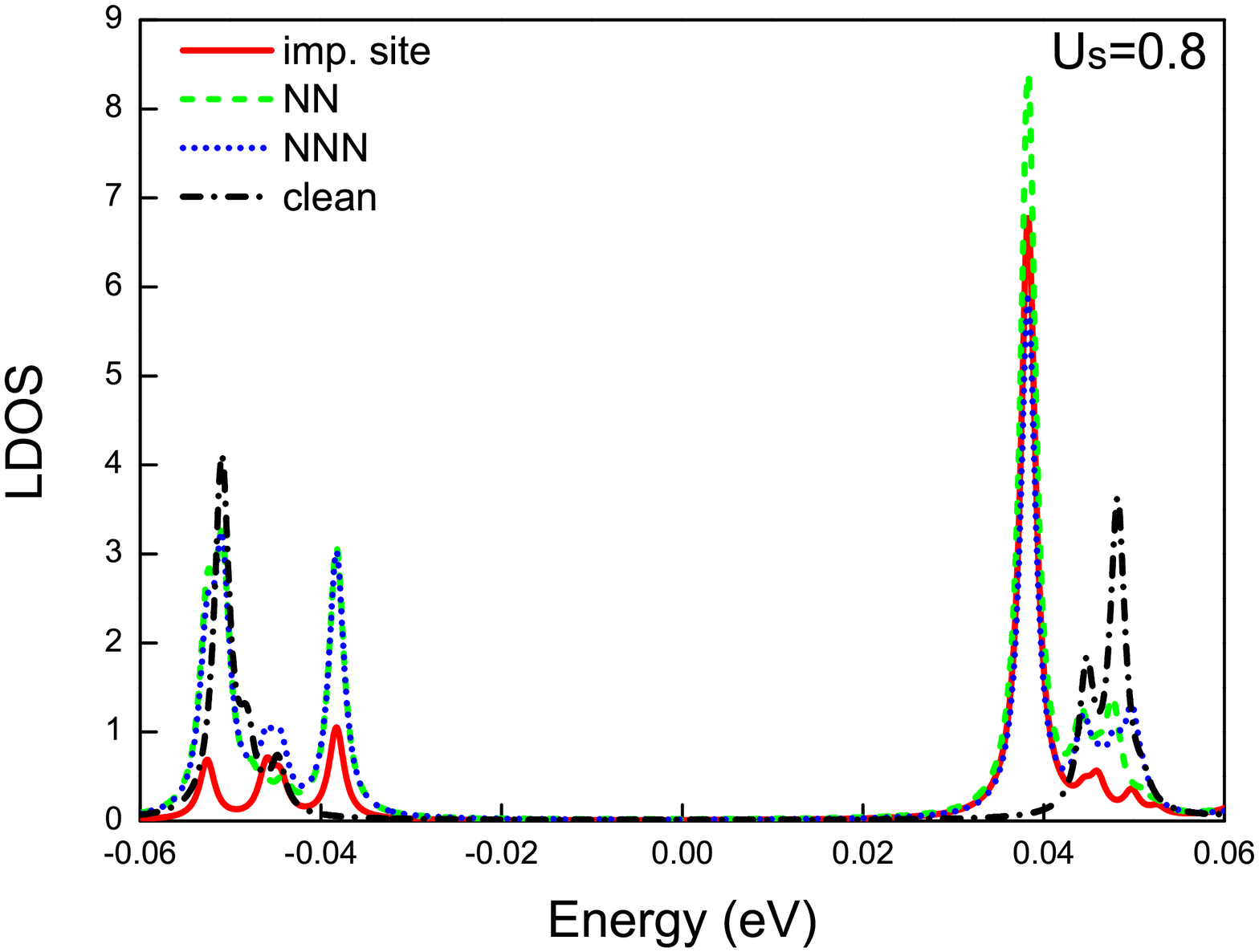}}
\subfigure[]{\includegraphics[width=120pt]{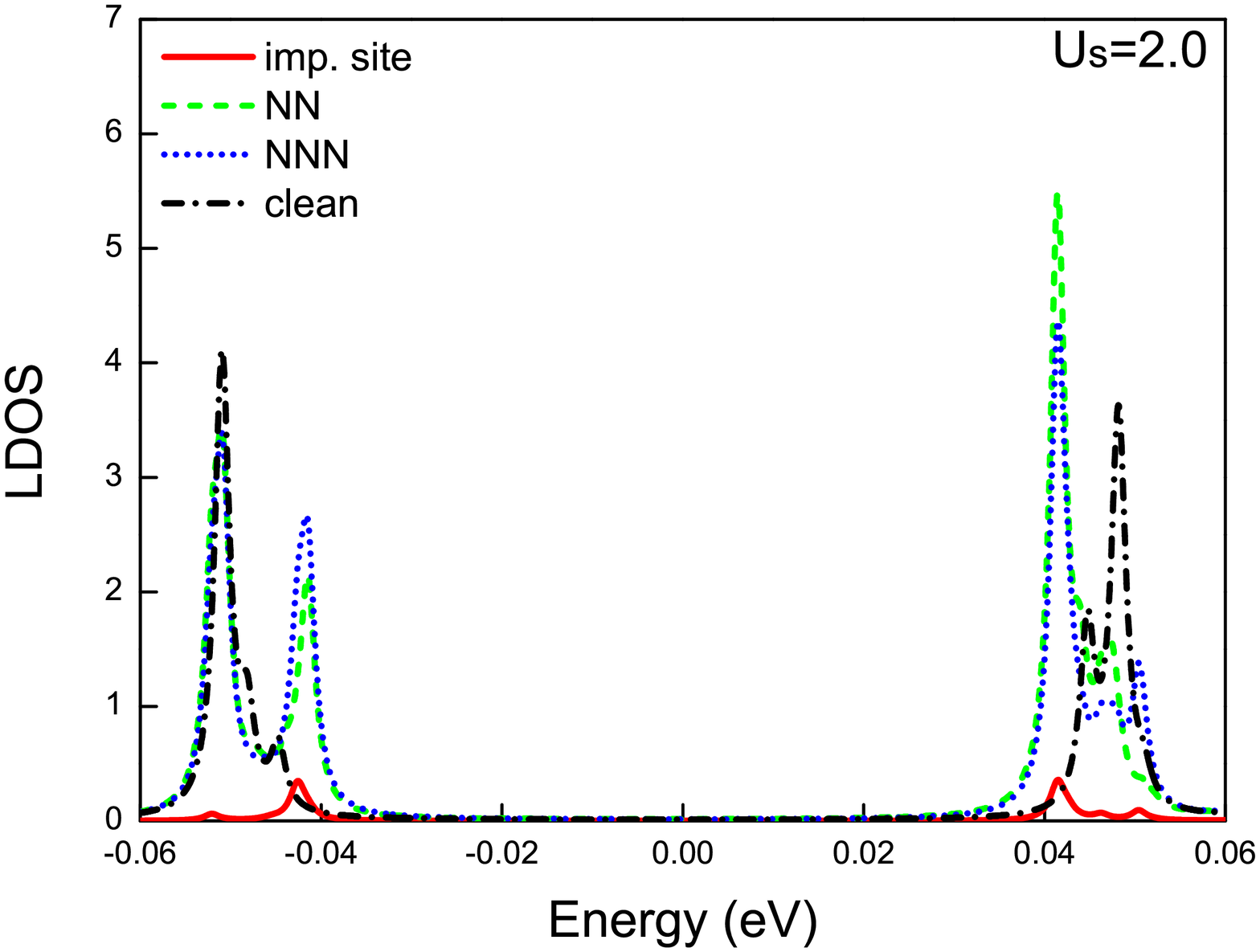}}
\subfigure[]{\includegraphics[width=120pt]{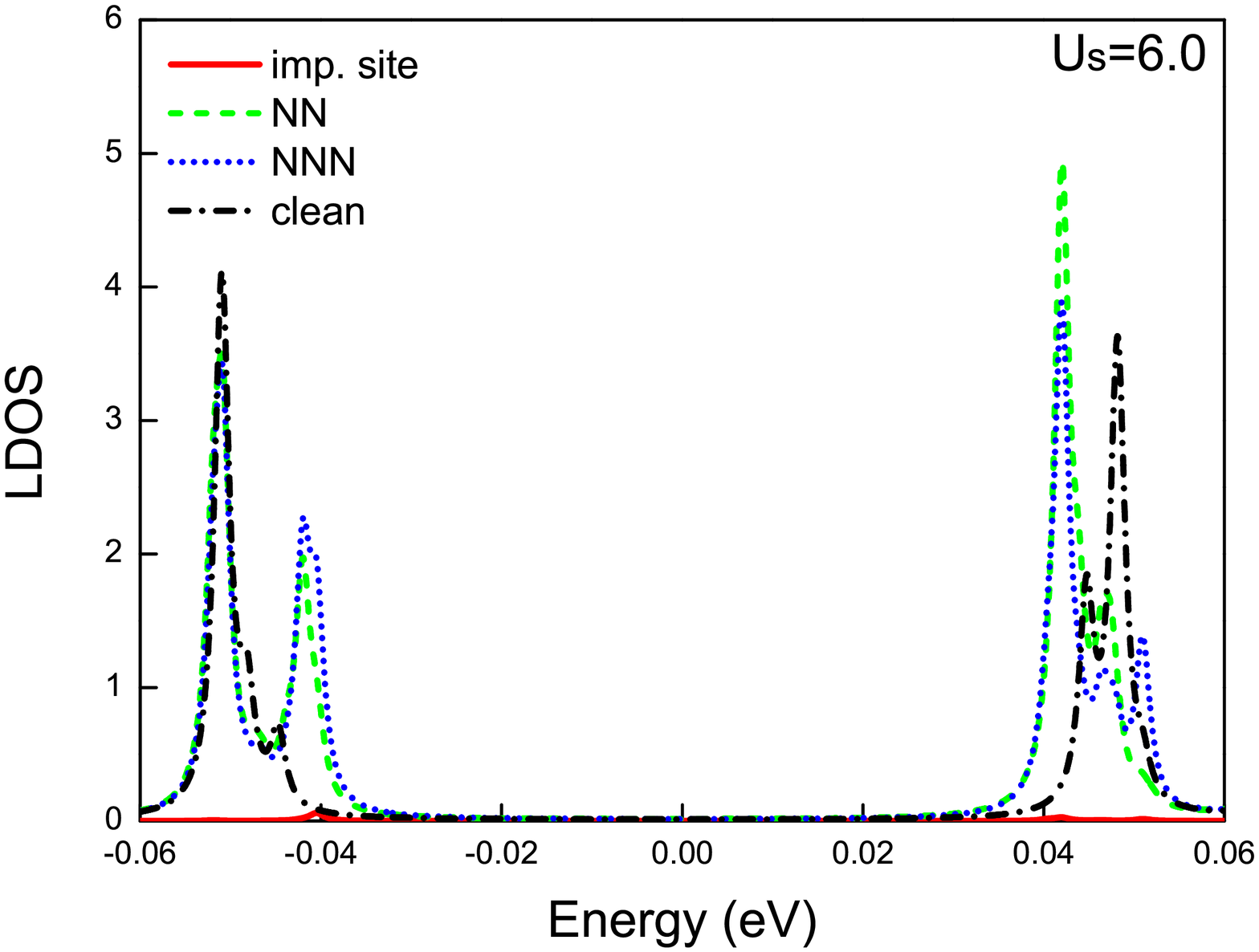}}
\caption{(color online) The LDOS for anisotropic $s$ wave pairing
state for repulsive scattering potential $U_{s}=0.5$ eV (a),
$U_{s}=0.8$ eV (b), $U_{s}=2.0$ eV (c), and $U_{s}=6.0$ eV (d). LDOS
for impurity, NN, and NNN sites are plotted in red(solid),
green(dashed), and blue(dotted) lines, respectively. The spectrum of
the clean system is also shown in black(dashed and dotted) line for
comparison.} \label{fig2}
\end{center}
\end{figure}

\begin{figure}[h]
\begin{center}
\subfigure[]{\includegraphics[width=120pt]{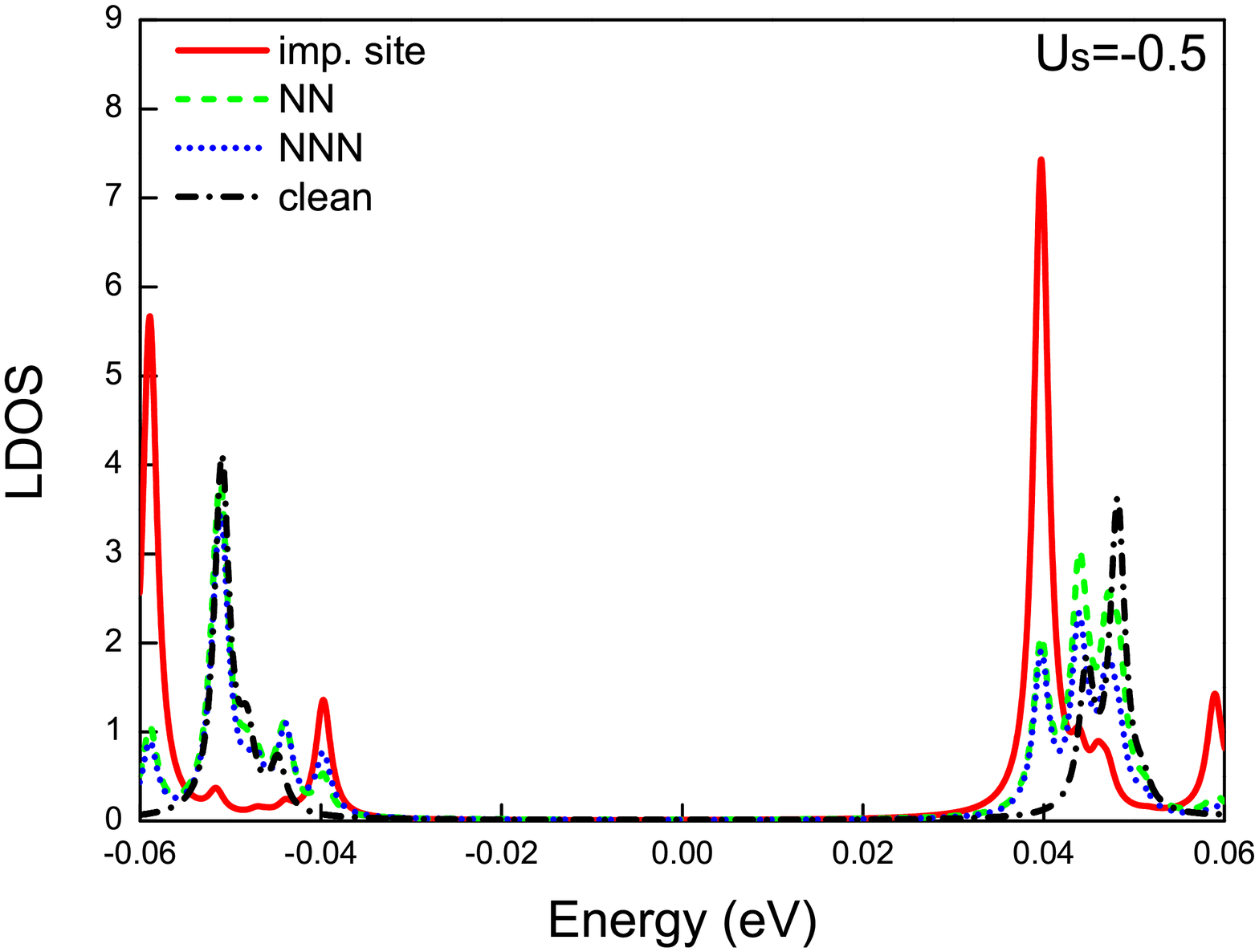}}
\subfigure[]{\includegraphics[width=120pt]{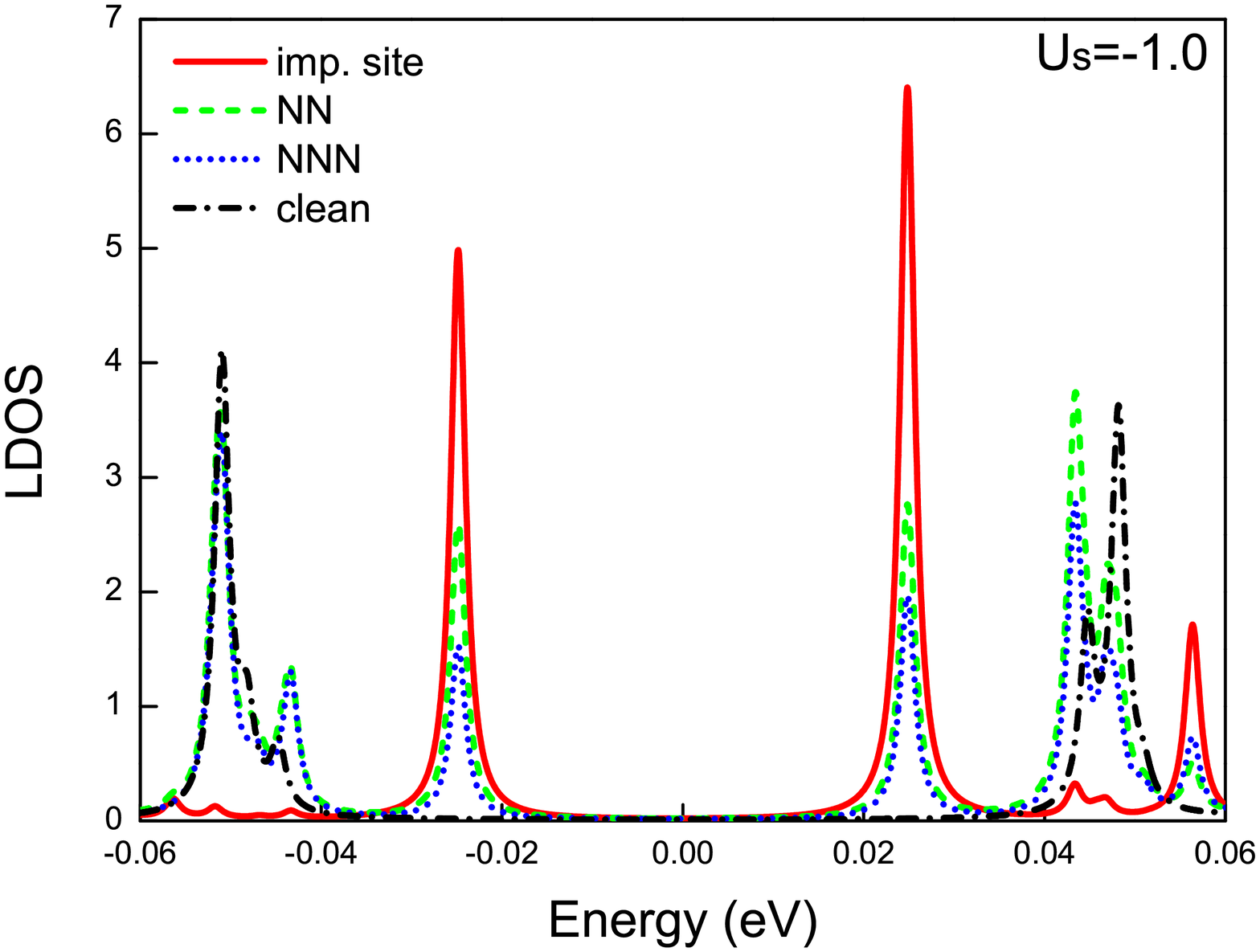}}
\subfigure[]{\includegraphics[width=120pt]{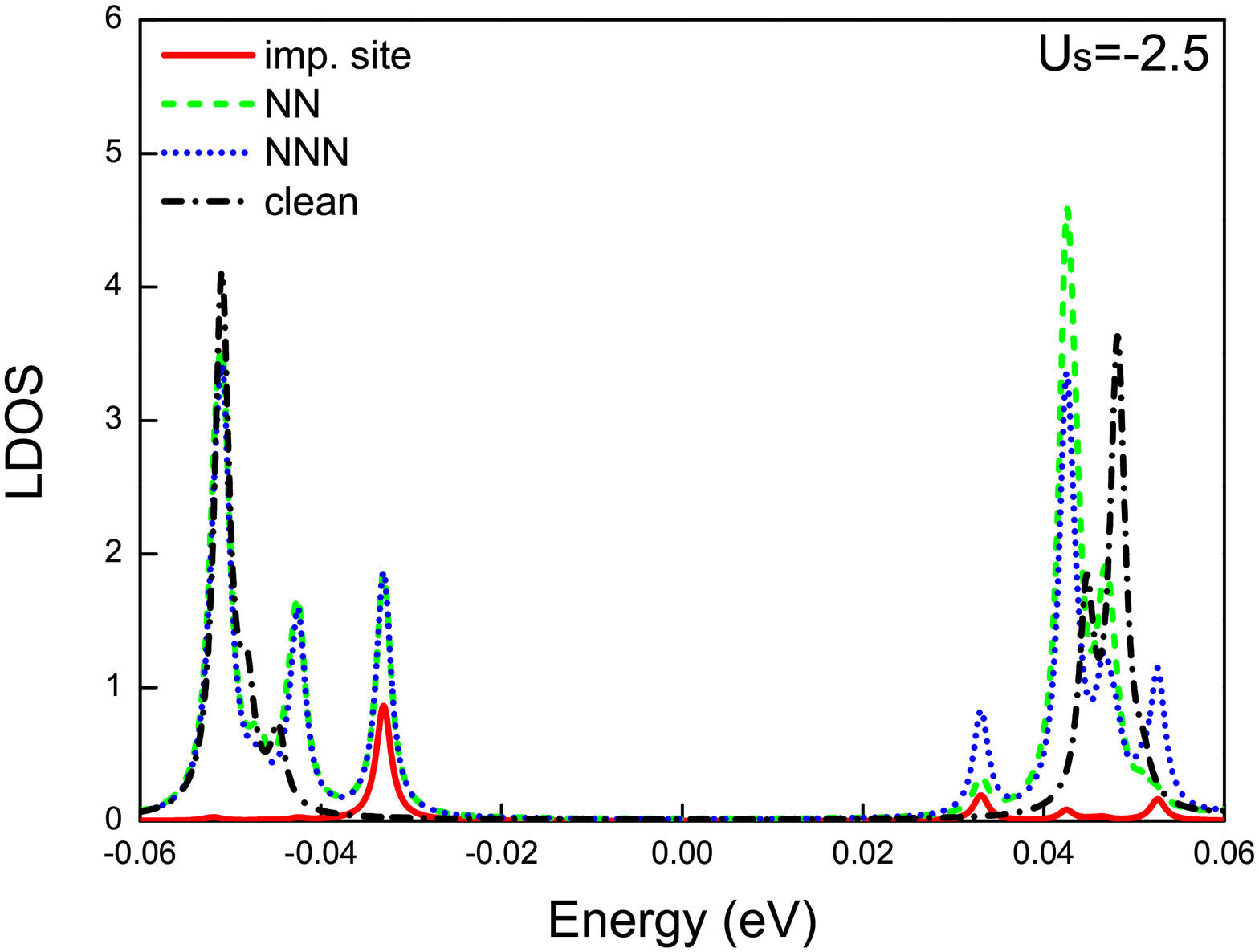}}
\subfigure[]{\includegraphics[width=120pt]{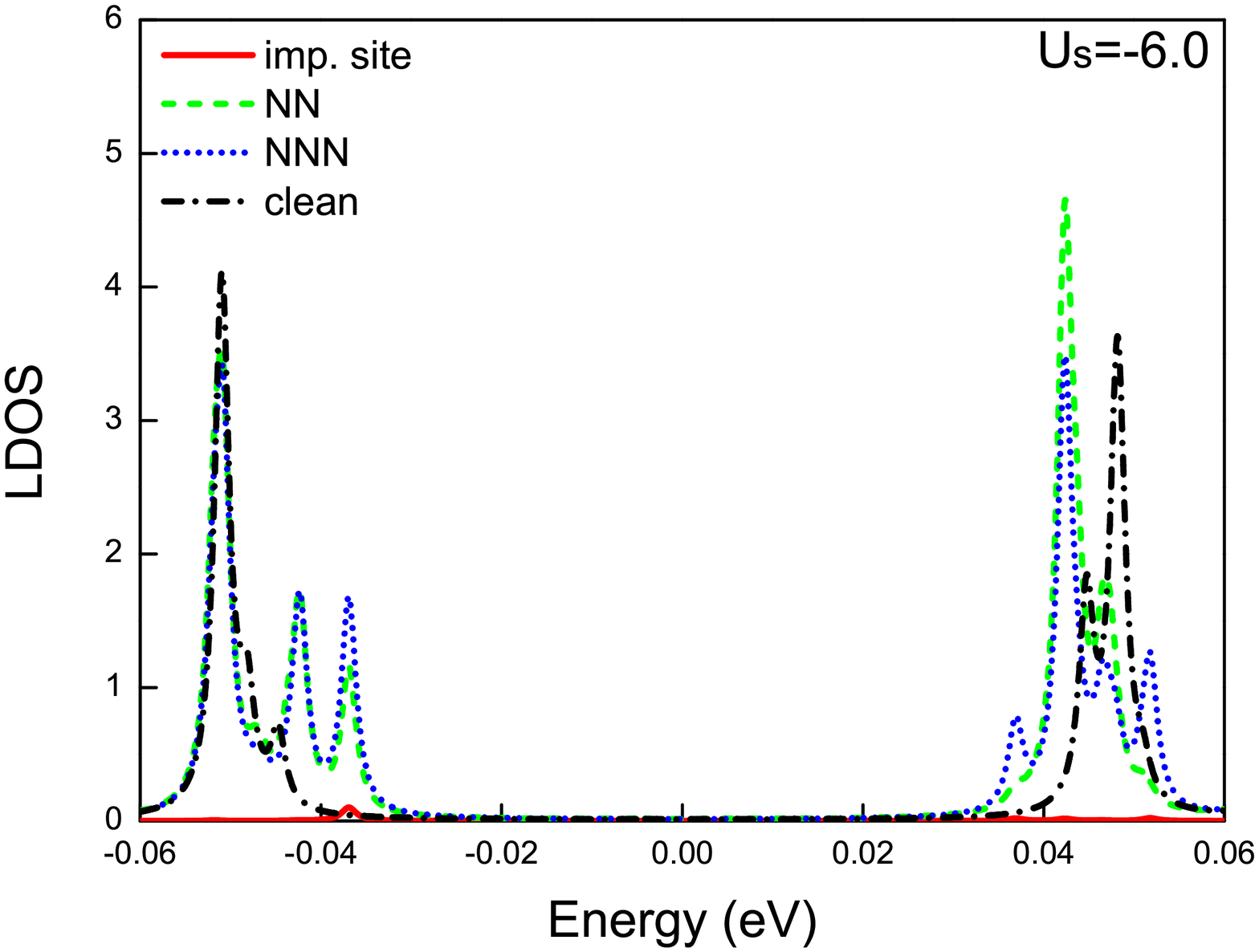}}
\caption{(color online) The LDOS for anisotropic $s$ wave pairing
state for attractive scattering potential $U_{s}=-0.5$ eV (a),
$U_{s}=-1.0$ eV (b), $U_{s}=-2.5$ eV (c), and $U_{s}=-6.0$ eV (d).}
\label{fig3}
\end{center}
\end{figure}

For isotropic $s$ wave pairing state, the existence of impurity does
not change much of the superconducting properties. Similar results
are also obtained in Ref. \cite{T.Kariyado,T.Zhou1}, which
demonstrates that there is no in-gap bound state for such a pairing
symmetry, whereas the order parameter on the impurity site are
suppressed evidently.

Results of anisotropic $s$ pairing symmetry are shown in Fig.
\ref{fig2} for repulsive scattering potentials and in Fig.
\ref{fig3} for attractive cases, respectively. The SC coherence peak
is located at $\pm$0.044 eV, and the typical resonating states
induced by the impurity appear for both cases at $\pm$0.028 eV for
$U_{s}=0.5$ eV and $\pm$0.025 eV for $U_{s}=-1.0$ eV, respectively.
The resonance peaks move towards the gap-edge as the amplitude of
the scattering potential increases for repulsive interaction, while
for attractive cases, the resonating peaks maintain at the gap-edge
for small value of scattering potential $U_{s}=-0.5$ eV, as shown in
Fig. \ref{fig3}(a). Additionally, it has been noticed that the
resonating states for positive interaction $U_{s}=0.5$ eV is very
robust in comparison with other values of $U_{s}$. At unitary limit,
the characteristics of the spectrum for both repulsive and
attractive interaction tends to be identical since the electrons
cannot hop to the impurity site, regardless of the detailed feature
of the scattering interaction. We note that the resonating states
within the SC gap, as compared with the clean cases, appear always
in pairs which are located symmetrically with respect to the Fermi
level. The magnitudes of the LDOS of the resonating states at
$E=\pm0.028$($\pm0.025$) eV for $U_{s}=0.5(-1.0)$ eV, for instance,
may not be the same, but they are of the same order. Therefore, we
may interpret them as the SC pairing states influenced by the
impurity.

\begin{figure}[!h]
\begin{center}
\subfigure[]{\includegraphics[width=120pt]{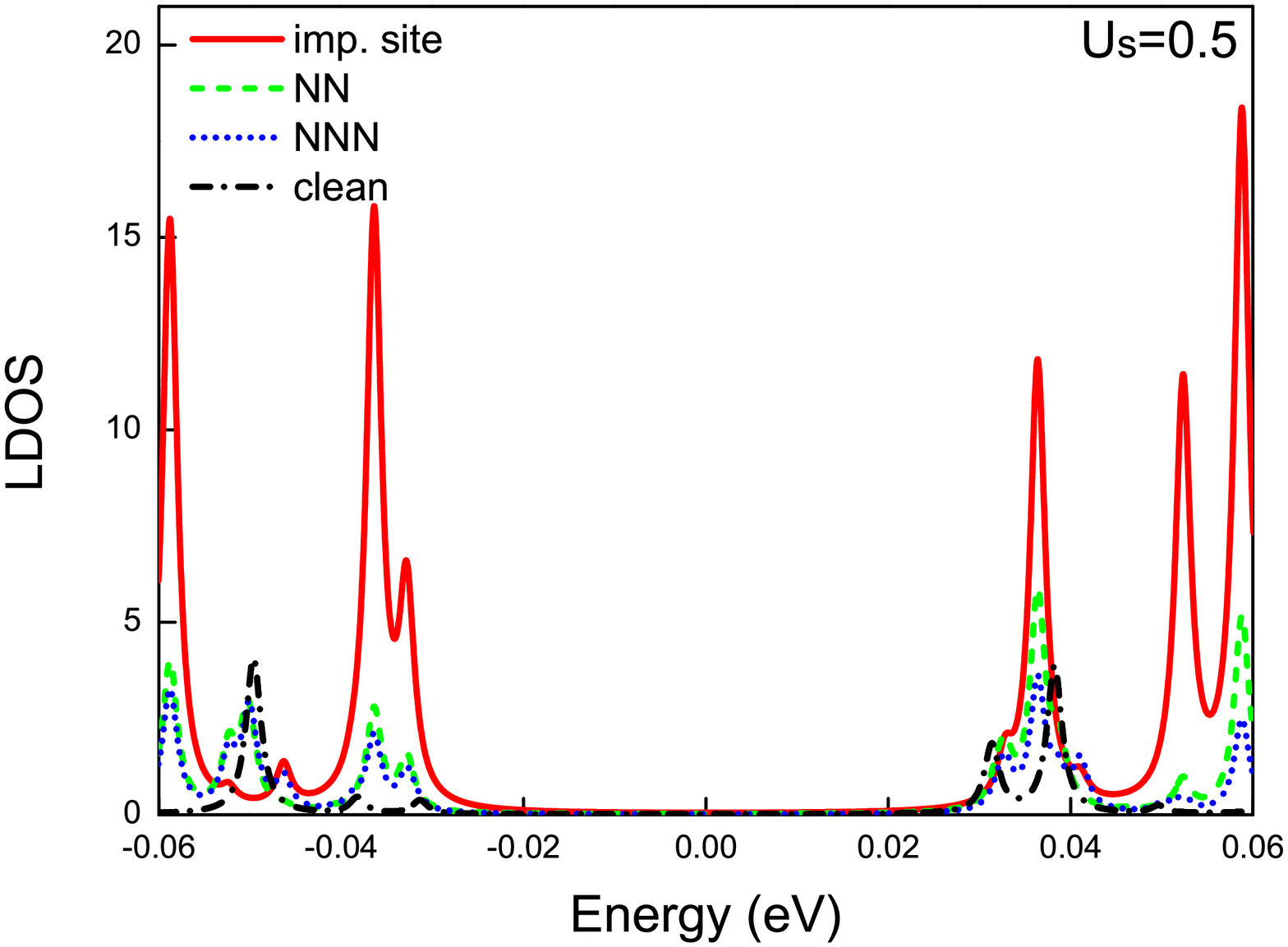}}
\subfigure[]{\includegraphics[width=120pt]{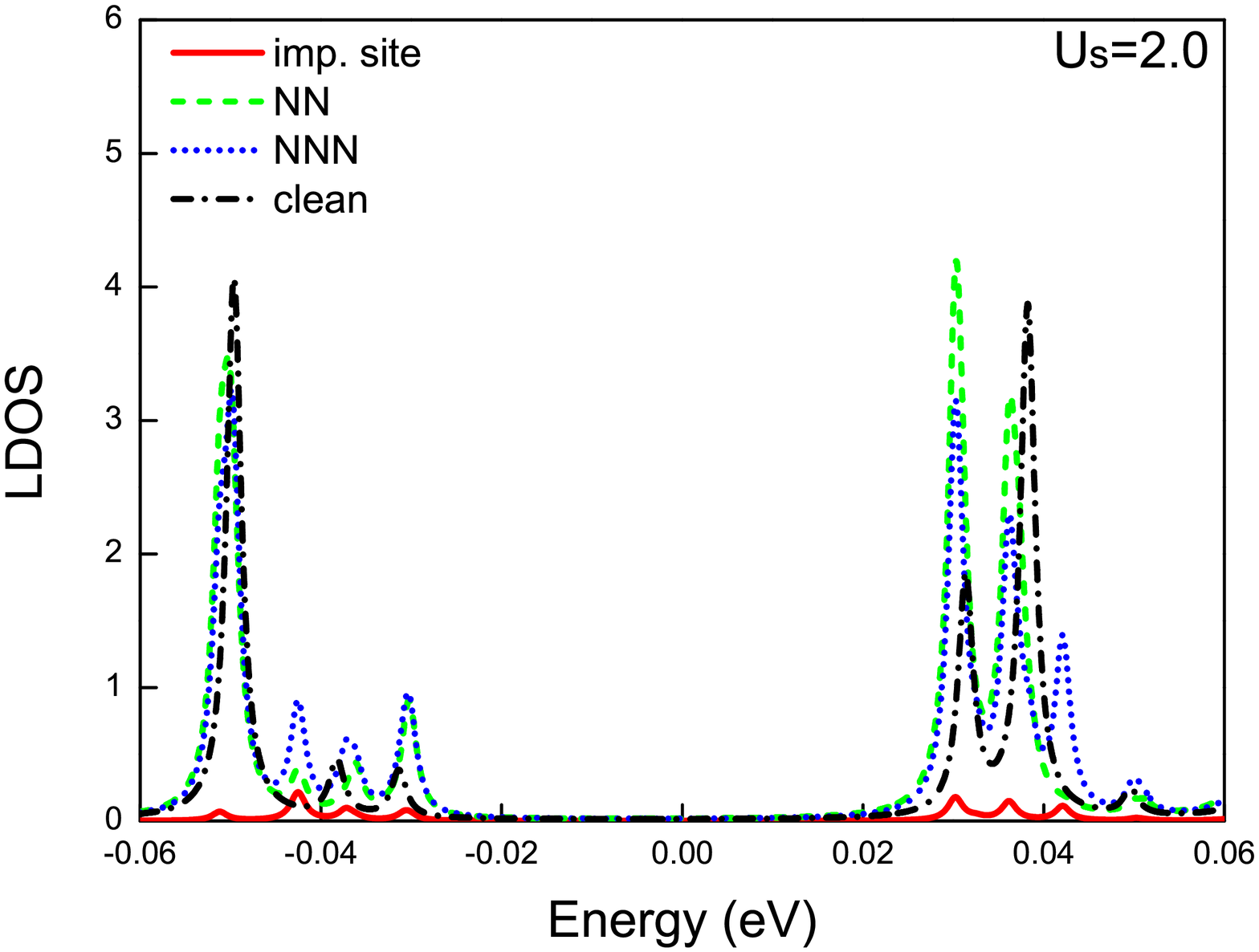}}
\caption{(color online) The LDOS for $d_{x^2-y^2}$ wave pairing
symmetry for repulsive scattering potential $U_{s}=0.5$ eV (a) and
$U_{s}=2.0$ eV (b).} \label{fig4}
\end{center}
\end{figure}

\begin{figure}[!h]
\begin{center}
\subfigure[]{\includegraphics[width=120pt]{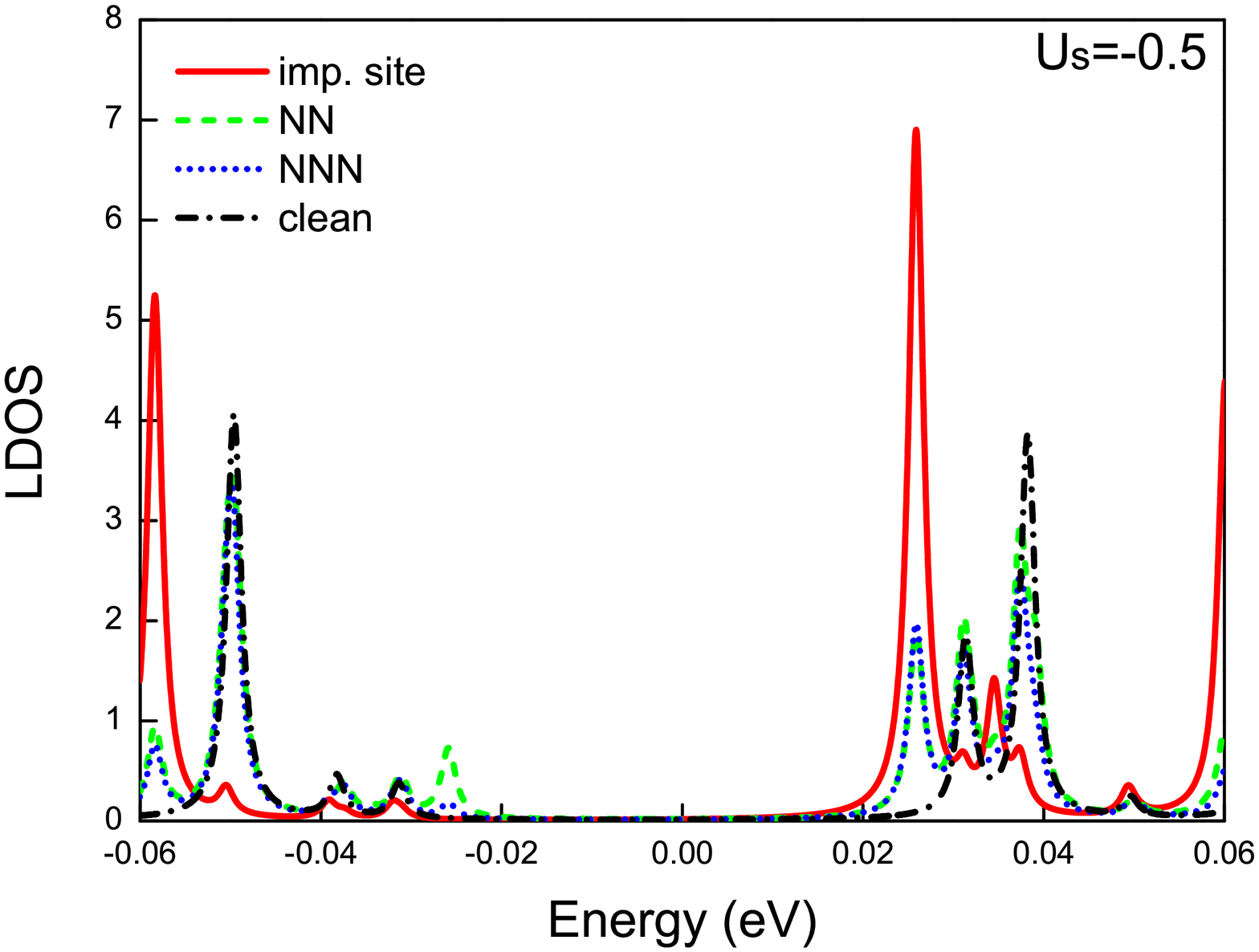}}
\subfigure[]{\includegraphics[width=120pt]{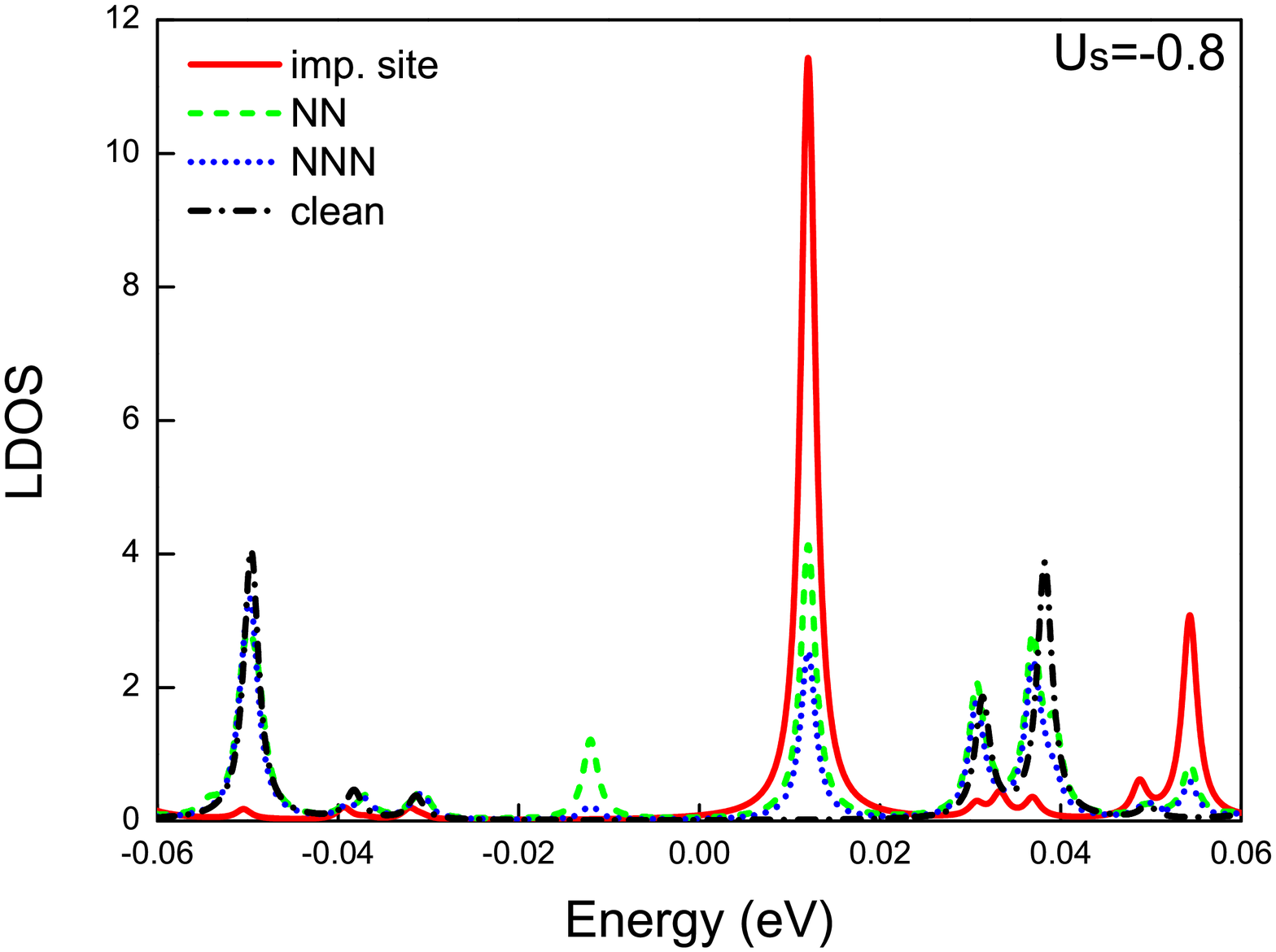}}
\subfigure[]{\includegraphics[width=120pt]{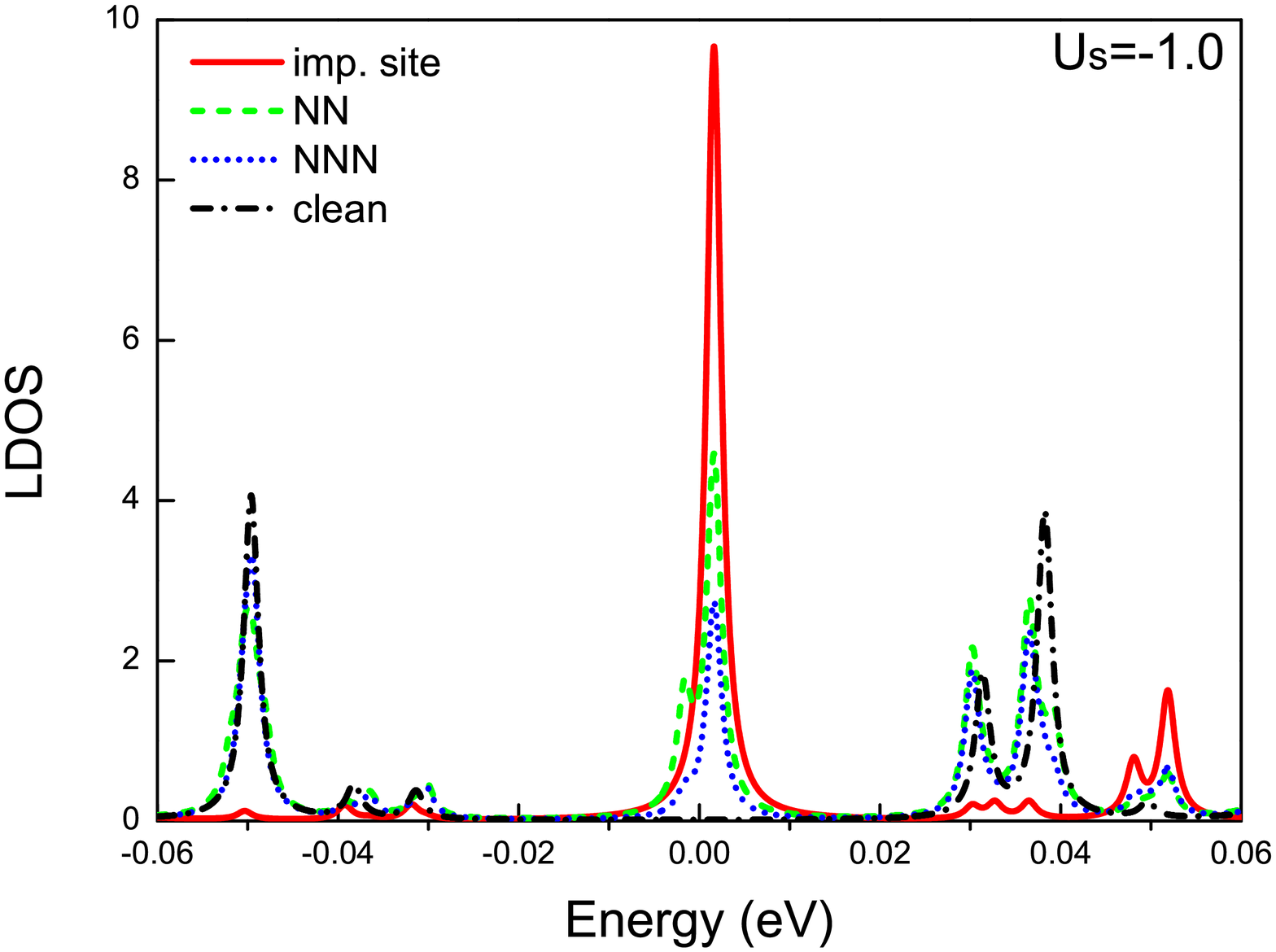}}
\subfigure[]{\includegraphics[width=120pt]{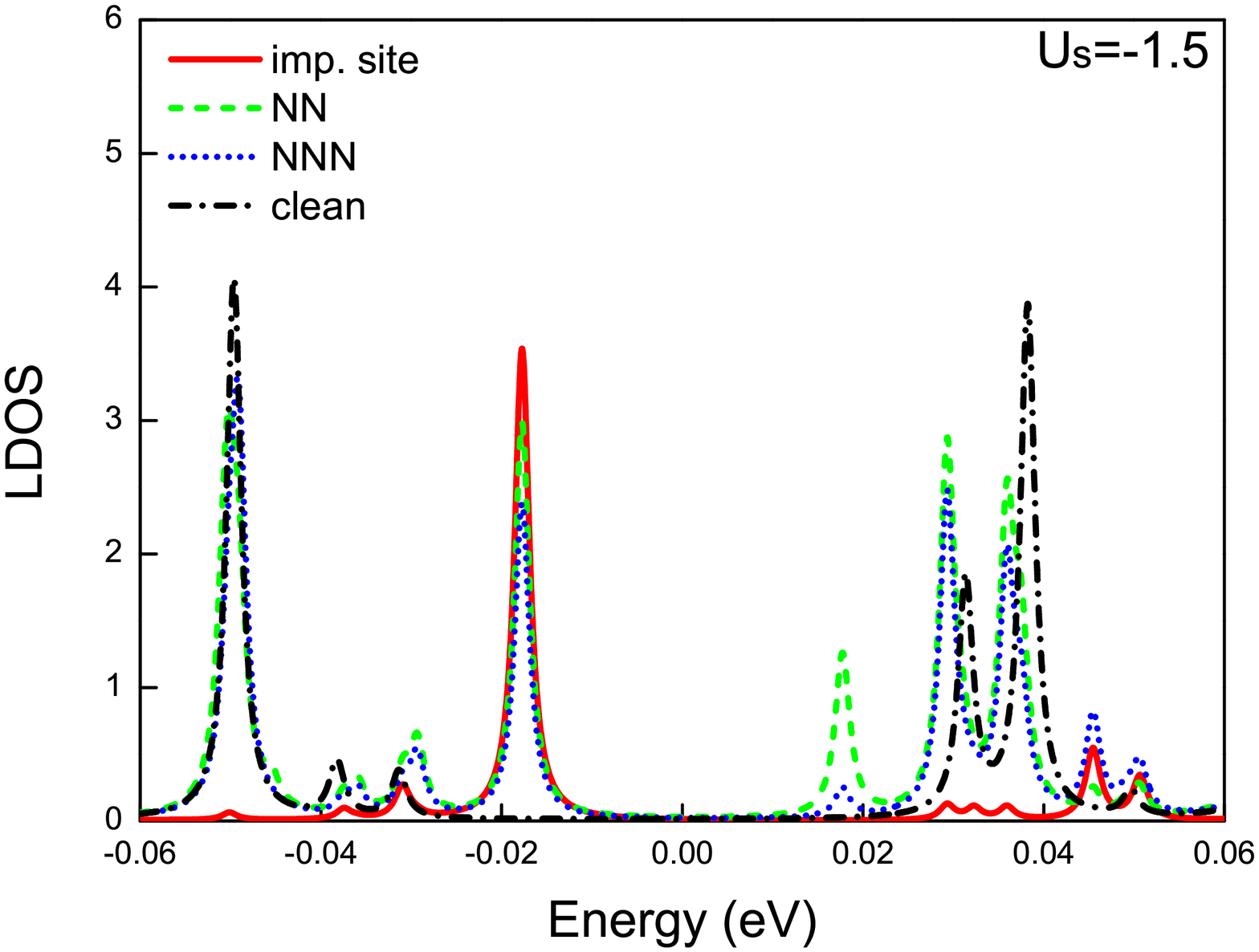}}
\subfigure[]{\includegraphics[width=120pt]{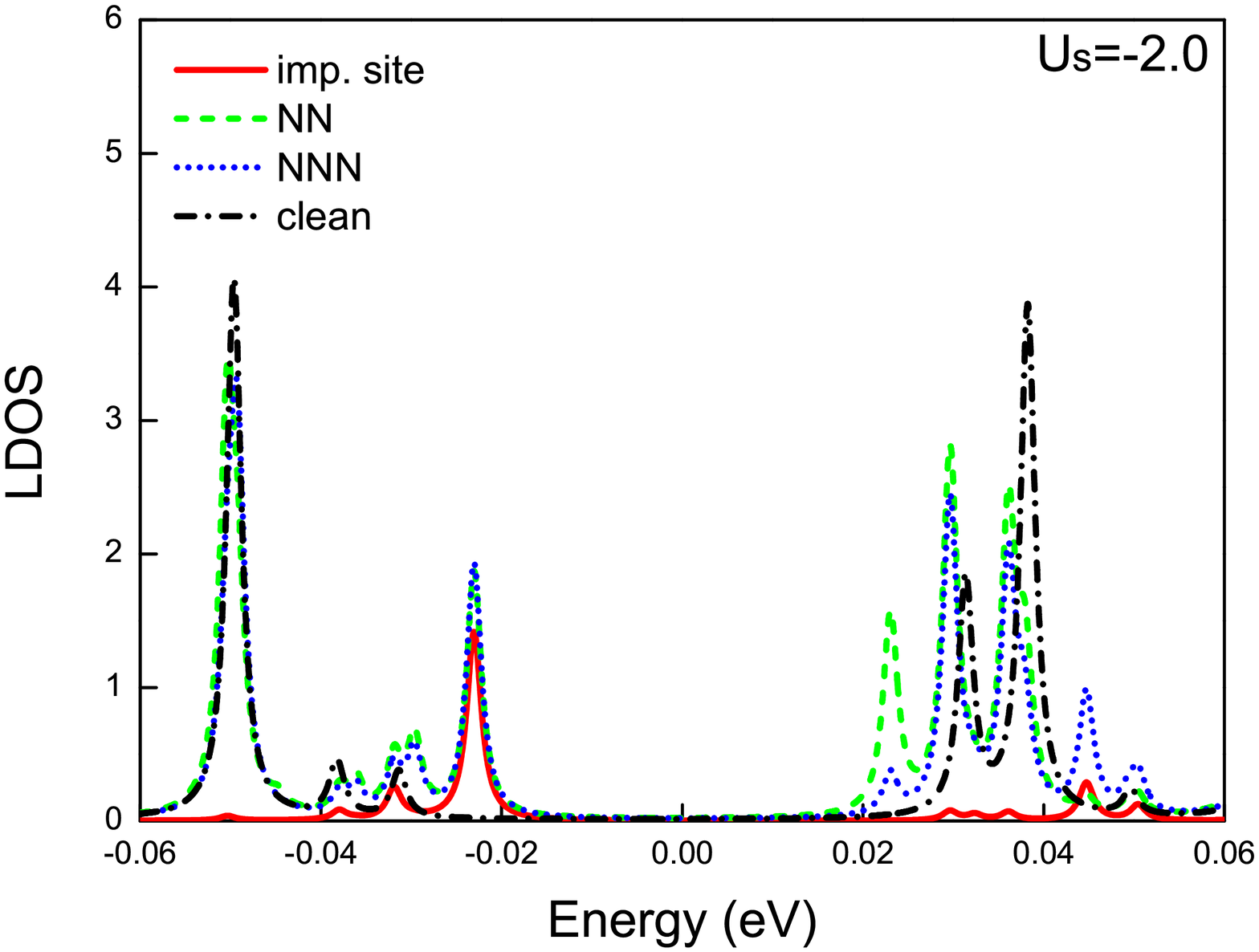}}
\subfigure[]{\includegraphics[width=120pt]{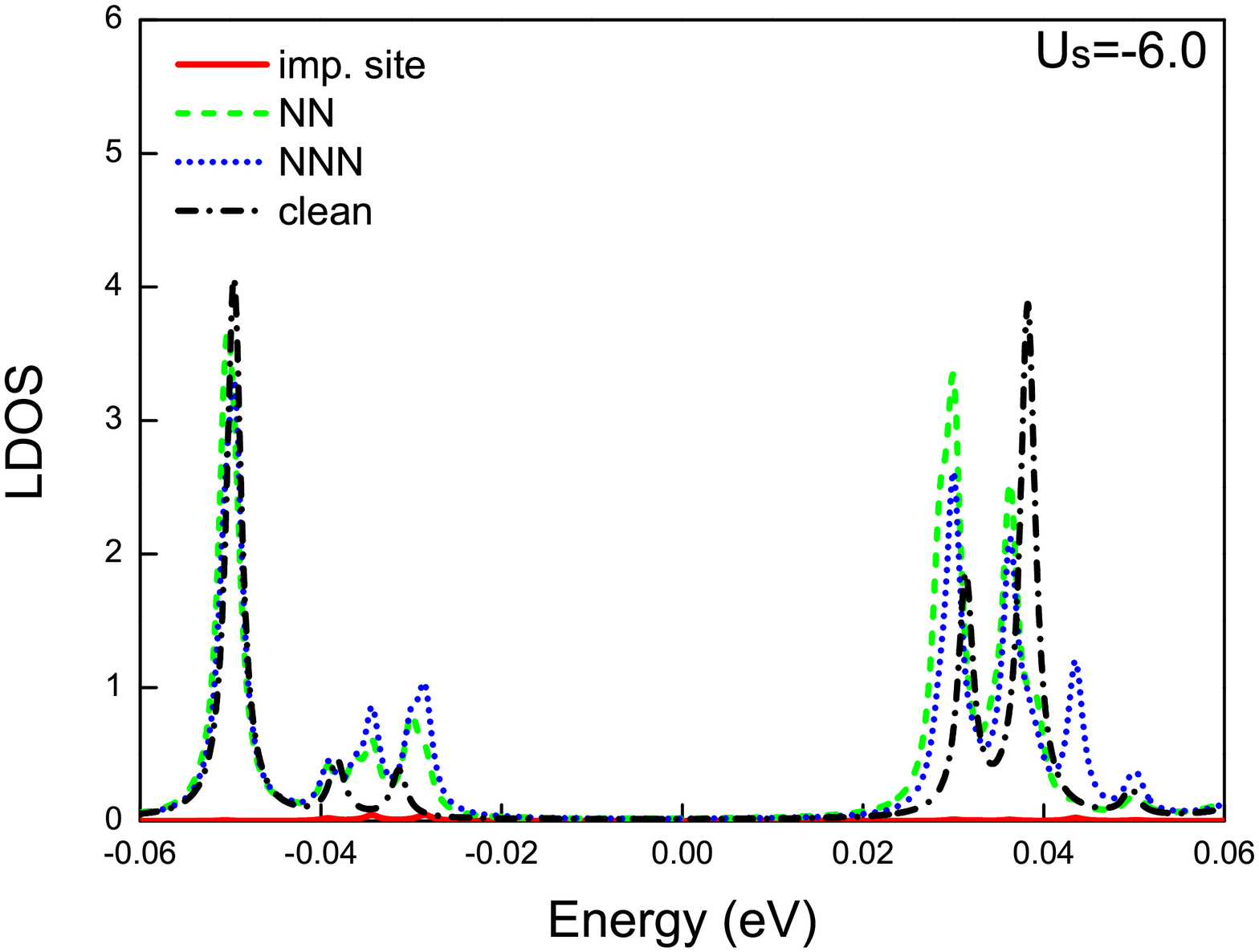}}
\caption{(color online) The LDOS for $d_{x^2-y2}$ wave pairing state
for attractive scattering potential $U_{s}=-0.5$ eV (a),
$U_{s}=-0.8$ eV (b), $U_{s}=-1.0$ eV (c), $U_{s}=-1.5$ eV (d),
$U_{s}=-2.0$ eV (e), and $U_{s}=-6.0$ eV (f).} \label{fig5}
\end{center}
\end{figure}

\begin{figure}[!h]
\begin{center}
\subfigure[]{\includegraphics[width=120pt]{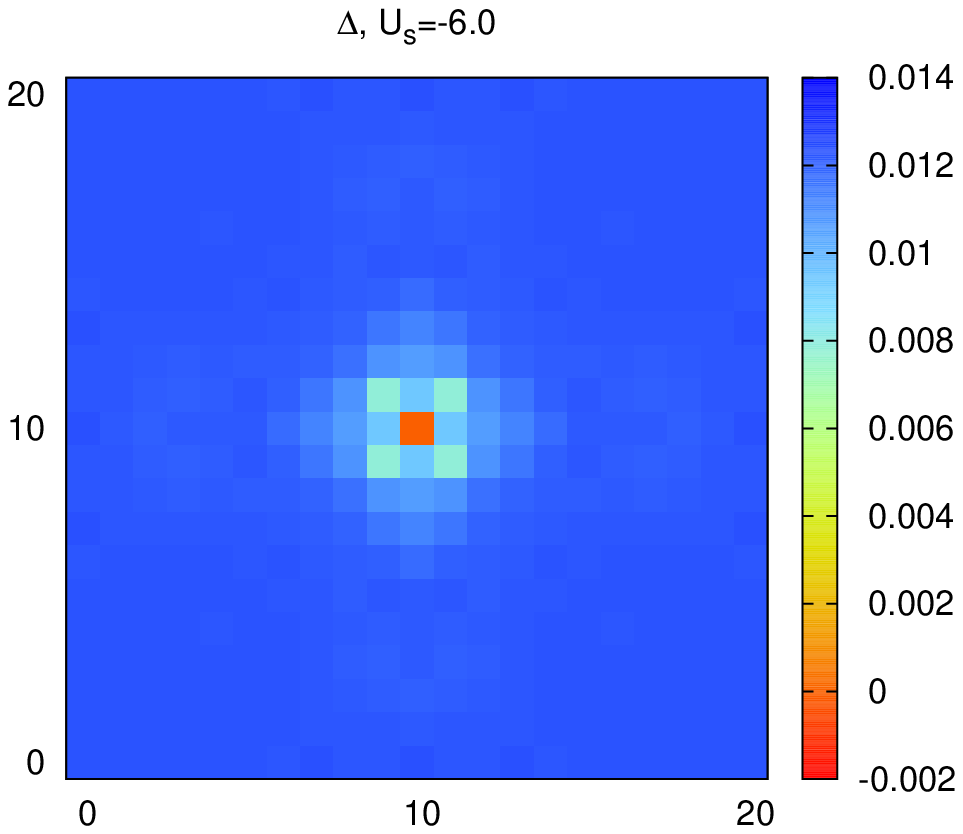}}
\subfigure[]{\includegraphics[width=120pt]{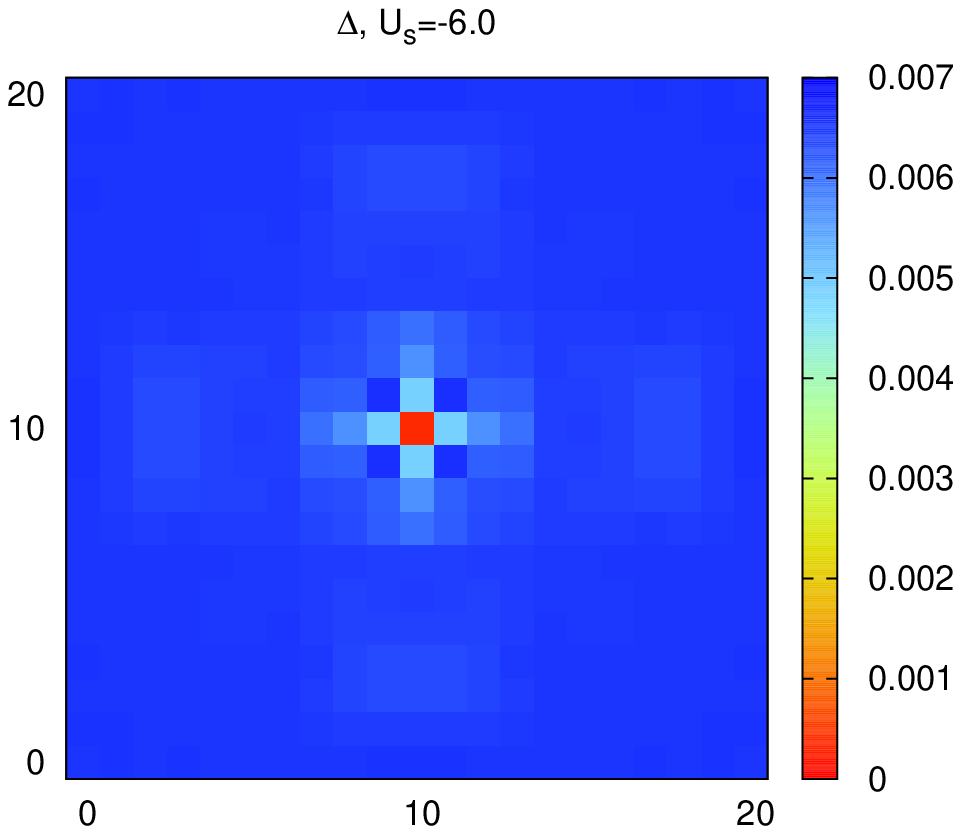}}
\caption{(color online) Real-space distribution of the order
parameter $\Delta(i)$ of anisotropic $s$ (a) and $d_{x^2-y^2}$ (b)
wave pairing states at unitary limit of the scattering interaction.}
\label{fig6}
\end{center}
\end{figure}

The response of $d_{x^2-y^2}$ wave pairing state to impurity
scattering in cuprate superconductors has been investigated
intensively, and the study of the impurity effects in
superconductors has been considered as an effective method to reveal
the pairing symmetry of the superconductivity \cite{A.V.Balatsky}.
In Fig. \ref{fig4} and \ref{fig5}, we show the LDOS for the
$d_{x^2-y^2}$ wave pairing state for positive and negative
scattering potentials, respectively. As we can see from the figures,
the in-gap bound states are only found for impurities with
attractive scattering potential, but not for impurities with
repulsive potential. As the impurity scattering potential increases
from -0.5 eV, the energy level of the impurity bound state shifts
continuously from well above the Fermi energy to the negative energy
then to the gap-edge as the intensity of the potential approaches to
the unitary limit. Obviously, these are impurity-induced bound
states stemming from the sign-change of the $d_{x^2-y^2}$ wave SC
gap. In our calculations, no evident in-gap bound states are
observed when the scattering potentials are small enough as
$|U_{s}|\leq0.4$ eV, which is physically justified that the
impurity-induced bound states would become small and vanish as the
amplitude of scattering interaction decreases. Decomposed LDOS
analysis demonstrates that the contribution of $d_{xy}$ orbital is
larger than that of $d_{xz}$ and $d_{yz}$ orbitals which indicates
the significance of the multi-orbital feature of Fe-based
superconductors. In passing, we note that the single
impurity-induced in-gap bound state, for the $d_{x^2-y^2}$ wave
pairing symmetry, is found near the Fermi energy for a moderate
scattering potential $U_{s}=-1.0$ eV. This is reminiscent of the
cuprates, where the emergence of the impurity-induced bound states,
as a result of the nodal $d_{x^2-y^2}$ wave pairing state, is
characterized by its low-lying feature.

We now discuss the asymmetry of the impurity states between
attractive and repulsive potentials. In a superconducting state, the
quasiparticle at the Fermi surface has exact electron-hole symmetry
in the sense it carries no net electric charge, $(|u(\vec q)| =
|v(\vec q)|)$. Away from the Fermi surface, the quasi-particle is
either more electron-like (outside the Fermi surface) or more
hole-like (within the Fermi surface).  In a usual superconducting
state, the gap function is approximately symmetric with respect to
outside or inside the Fermi surface. Therefore, quasi-particles
outside or within the Fermi surface have similar energies, and there
is an approximate electron-hole symmetry in responding to a charged
impurity. Since a repulsive potential to an electron means an
attractive potential to a hole, we expect a symmetry in the impurity
bound state between attractive and repulsive potentials with respect
to an electron.  This is what we usually also find in study of the
impurity bound state.  The situation in a FeSe-based superconductor
is however, very different. As we can see from Fig. 1, the gap
$\Delta(\vec k)$ vanishes along the zone diagonal, well outside the
electron Fermi surfaces. Therefore, electron-like quasi-particles
(outside the Fermi surface) contributes more to the impurity state
than the hole-like quasiparticles (inside the Fermi surface), and
there is lack of electron-hole symmetry in the electronic structure.
Our results for existence of the bound impurity state for attractive
potential but not for repulsive potential may be understood as the
strong electron-hole asymmetry. Mathematically, the amplitude of the
density of the impurity-induced bound states is determined by the
Bogoliubov quasiparticle weights $|u^{n}_{i\alpha}|^2$ and
$|v^{n}_{i\alpha}|^2$, and the asymmetry of LDOS of the bound states
in Fig. \ref{fig5} demonstrates a strong violation of the
particle-hole symmetry. We note that using $T$-matrix method, the
impurity-induced bound states in (K,Tl)Fe$_{x}$Se$_{2}$
superconductors have been predicted by Zhu \emph{et
al.}\cite{J.X.Zhu} in the case of $d_{x^2-y^2}$ pairing state. Our
results, which are obtained from a microscopic orbital-featured
Hamiltonian, are qualitatively similar to the findings in the
previous work \cite{J.X.Zhu2} except that additional information has
been revealed that the emergence of the impurity-induced bound state
requires an attractive on-site scattering interaction which
originates from the effective potential difference between the
impurity-3d and Fe-3d $t_{2g}$ electrons.

In Fig. \ref{fig6}, we plot the distribution of the order parameter
in real-space at unitary limit of the impurity scattering. It shows
a color mapping of the cross shaped \cite{A.V.Balatsky} suppression
of the order parameters which occurs prominently along the NNN
directions for anisotropic $s$ wave and NN directions for
$d_{x^2-y^2}$ wave pairing state, respectively.

Additionally, the subtle mechanism of the substitution of Co and Ni
for Fe remains a controversial question in Fe-based superconductors.
The effective impurity potential induced by various dopants which
can be ascribed to the difference of the on-site ionic potential,
together with the modification of the Fermi surface caused by the
rigid shift of the band structure, has been investigated from both
theoretical and experimental perspective
\cite{K.Nakamura,H.Wadati,G.Levy}. The isovalence of Co and Fe in
Ca(Fe$_{0.944}$Co$_{0.056}$)$_{2}$As$_{2}$ compound has been pointed
out and a binding energy shift (0.25 eV) has also been observed,
which reveal the complexity of the impurity nature of Co
\cite{G.Levy}. Inspired by this finding, one may anticipate that the
doping of Ni in iron pnictide or chalcogenide superconductors may be
another promising alternative since the effective impurity
potential, as reported by Nakamura \emph{et al.} \cite{K.Nakamura},
is higher than that of Co. The impurity-induced bound states
observed in our calculations in the state of $d_{x^2-y^2}$ wave
pairing when the scattering potential is within the range of
$U_{s}=-0.5\sim-6.0$ eV may simulate the cases of Ni and Co doping
in FeSe-based 122-type superconductors.

In summary, using a three-orbital model, we present a comprehensive
investigation of the effects of a single nonmagnetic impurity in
A$_{x}$Fe$_{2-y}$Se$_{2}$ (A=K, Rb, or Cs) superconductors by means
of BdG theory. The in-gap bound states have been found for
$d_{x^2-y^2}$ wave pairing state. For anisotropic $s$ wave pairing
state, the observed resonating states are typically SC pairing
states influenced by impurity. The locations of the bound states for
$d_{x^2-y^2}$ pairing symmetry, moving towards the Fermi energy from
the gap-edge as the amplitude of the scattering potential increases
from 0.5 to 1.0 (eV), appear uniquely for attractive interaction.
Due to the absence of the hole pocket, strong violation of the
electron-hole symmetry occurs, which results in that no
impurity-induced bound state emerges for repulsive scattering
interaction. Therefore, distinguishable responses to a single
nonmagnetic impurity, in terms of LDOS on and near the impurity
site, would be an approach to examine the pairing symmetries of
FeSe-based 122-type superconductors.

\acknowledgments We acknowledge financial support from Hong Kong RGC
HKU 706809 and NSFC 11274269.

\end{document}